# Photon Walk in Transparent Wood: Scattering and Absorption in Hierarchically Structured Materials

*Hui Chen, Céline Montanari, Ravi Shanker, Saulius Marcinkevicius, Lars A. Berglund, and Ilya Sychugov\**

H. Chen, C. Montanari, R. Shanker, L. A. Berglund
Wallenberg Wood Science Center, Department of Fiber and Polymer Technology, KTH Royal Institute of Technology, Teknikringen 56, 100 44 Stockholm, Sweden

S. Marcinkevicius, I. Sychugov
Department of Applied Physics, School of Engineering Sciences, KTH Royal Institute of Technology, Hannes Alfvens väg 12, 114 19 Stockholm, Sweden
E-mail: ilyas@kth.se

The optical response of hierarchical materials is convoluted, which hinders their direct study and property control. Transparent wood (TW) is an emerging biocomposite in this category, which adds optical function to the structural properties of wood. Nano- and microscale inhomogeneities in composition, structure and at interfaces strongly affect light transmission and haze. While interface manipulation can tailor TW properties, the realization of optically clear wood requires detailed understanding of light-TW interaction mechanisms. Here we show how material scattering and absorption coefficients can be extracted from a combination of experimental spectroscopic measurements and a photon diffusion model. Contributions from different length scales can thus be deciphered and quantified. It is shown that forward scattering dominates haze in TW, primarily caused by refractive index mismatch between the wood substrate and the polymer phase. Rayleigh scattering from the wood cell wall and absorption from residual lignin have minor effects on transmittance, but the former affects haze. Results provide guidance for material design of transparent hierarchical composites towards desired optical functionality; we demonstrate experimentally how transmittance and haze of TW can be controlled over a broad range.

# 1. Introduction

Turbid media, from aerosols[1] to biological tissue,[2,3] is a class of materials where analysis of complex light interaction with matter can be used to understand its properties. Optically transparent wood (TW) is a new turbid material that combines high visible light transmittance with good mechanical properties,[4] featuring anisotropic scattering strength.[5] TW is normally prepared by infiltrating a refractive index (RI) matched polymer into a delignified wood substrate, so that the wood pore space is filled by a polymer phase. TW as a new concept has recently inspired numerous applications for functional biocomposites, such as in smart windows,[6] for light trapping on solar cells,[7,8] as a scaffold for nanoparticles,[9–11] for heat storage based on phase change fillers, and many others.[12,13]

Understanding the light-TW interaction is essential for further development and application of this material. In particular, controlling the haze (scattering strength) is important to realize TW tailored for applications where one may need either a high or a low extent of diffused light. The scattering mechanisms are related to its hierarchical structure, where both micro- and nanoscale features are present. An additional complication is that the spatial structure of wood is neither periodic nor purely random. Thus the very concept of size-independent average quantities, such as scattering and absorption coefficients, may not be applicable to TW, while those are widely used in studying turbid air,[14,15] sea water,[16,17] plastics,[18] as well as glass and its composites.[19,20]

In general, light interaction with a transparent composite material is influenced by (i) RI matching of different components, (ii) their structure and the composition. In the initial concept of TW,[21] the RI matching is realized through polymer mixtures. Thus, accurate data for the RI value of the wood substrate facilitates TW property tuning through RI matching of the polymer matrix phase. Recently, we directly measured the RI value of delignified wood (DW) using immersion liquid method, while accounting for the anisotropic wood structure. The RI of the DW perpendicular to the wood fiber direction was between 1.535 and 1.538 at a wavelength of 589 nm.[22] As for composition effects, we found that with increased lignin content the total transmittance decreased,[23] and also with increased cellulose volume fraction.[24] As a multiphase material, the wood phase in TW contains cellulose, hemicellulose and residual lignin and there is also a polymer phase filling the pore space. The exact composition and structure vary for different wood species, pre-treatments or cut directions (i.e. different volume fraction of cellulose or lignin content). In addition, there may be voids inside the sample (e.g. nano-voids inside wood cell wall or air gaps between cell wall and polymer) from inherent structure or TW preparation. Altogether it is challenging to identify and decouple different effects from each other, making adequate comparison and analysis of TW transmittance and haze data difficult.

Previously, attempts were made to describe light propagation in single wood fibers and paper by combining simulations, such as Monte Carlo, and ray-optics.[25,26] Granberg et al. has investigated the anisotropic scattering of fiber-containing surfaces. They have shown that the fiber orientation and the reflection/refraction from its inner and outer surfaces contribute to the anisotropic distribution of reflected light.[27,28] Linder et al. also found similar results pointing out to the alignment and conical scattering by the cylindrical wood fibers as the main source of the anisotropic scattering in paper.[25,29] Photon diffusion theory is another method to describe the photon propagation in highly scattering materials.[2,30–33] Kienle et al. has combined Monte Carlo simulations and diffusion theory to investigate the light propagation in native wood,[34] while TW optical response has not yet been thoroughly examined. Recently, Nishiyama summarized possible scattering processes in a nanocellulose film and TW without quantifying respective contributions.[35]

In this article, TW is described using sample size-independent scattering and absorption coefficients. We present a noninvasive and nondestructive approach to extract those for TW using photon budget measurements, and this methodology is suitable for a large class of materials. It is based on different angle distributions of the scattered light by nano- and microscale inhomogeneities. Here, by changing the TW preparation method, we established quantitative contributions from scattering processes of different origin. Experimental data and the developed light diffusion model are in support of forward scattering as a dominant mechanism in TW biocomposites, rather than Rayleigh scattering and light absorption. We find a concomitant increase of the latter two coefficients with higher cellulose volume fraction (lignin), pointing towards the important role of the cell wall. Based on the results we suggest materials design strategies to tune TW haze and transmittance towards optically clear wood.

## 2. Results and Discussion

### 2.1. Light Interaction with TW

Wood fibers are nearly cylindrical, irregular tubular cells, and consist of a cell wall (few microns in thickness) and an empty central lumen space (tens of microns in diameter), see **Figure 1**a. The cell wall itself is highly porous and is made of aligned cellulose nano-fibrils in a lignin/hemicellulose matrix, manifesting a hierarchical wood structure.[36] In TW biocomposites, a RI-matched polymer matrix is added to fill micro- and nanoscale pores of the DW substrate. As a result, an optically homogeneous medium is obtained so that light-scattering is reduced. In this study, we prepared TW samples from balsa wood with various densities, varying in sample thickness and volume fraction of cellulose ($V_f$). The main difference between the wood specimens is cell wall thickness and lumen diameter, as shown schematically in Figure 1a. In addition, we probed three different fabrication methods to control the interface between the wood cell wall and the polymer: DW infiltrated with pre-polymerized methyl methacrylate (PMMA, RI = 1.49), named as DW-PMMA; succinic anhydride (SA) treated DW (RI = 1.533, measured based on our previous method[22] as shown in Figure S1a) infiltrated with PMMA, named SA-PMMA; and, finally, SA-treated DW infiltrated with poly(limonene acrylate) with a more suitable RI (PLIMA, RI = 1.52[37]), referred to as SA-PLIMA.

The interface between cell wall and lumen polymer will vary depending on the nature of the wood-polymer interaction. When wood-polymer adhesion is weak, the presence of micro- and nanoscale air gaps can be identified at the cell wall-lumen interface.[38] For example, DW-PMMA has weak wood-polymer interaction as shown in Figure 1b, resulting in debonding gaps after polymerization-related shrinkage. For SA-PMMA samples, improved molecular interactions are achieved due to the SA wood substrate modification providing stronger cell wall-polymer adhesion. The role of SA treatment is to reduce moisture content in the wood substrate and to facilitate monomer diffusion in the cell wall.[39] Finally, SA-PLIMA stands out since covalent bonds are formed between polymer and cell wall.[37] The transparency of DW-PMMA, SA-PMMA and SA-PLIMA is illustrated in the photographs in Figure 1c, where the TW samples ($V_f$ = 4.5%, thickness = 0.2 cm) are on the top of an "@" symbol. For the TW composite from SA-PLIMA, the symbol is the clearest, which means scattering is the weakest.

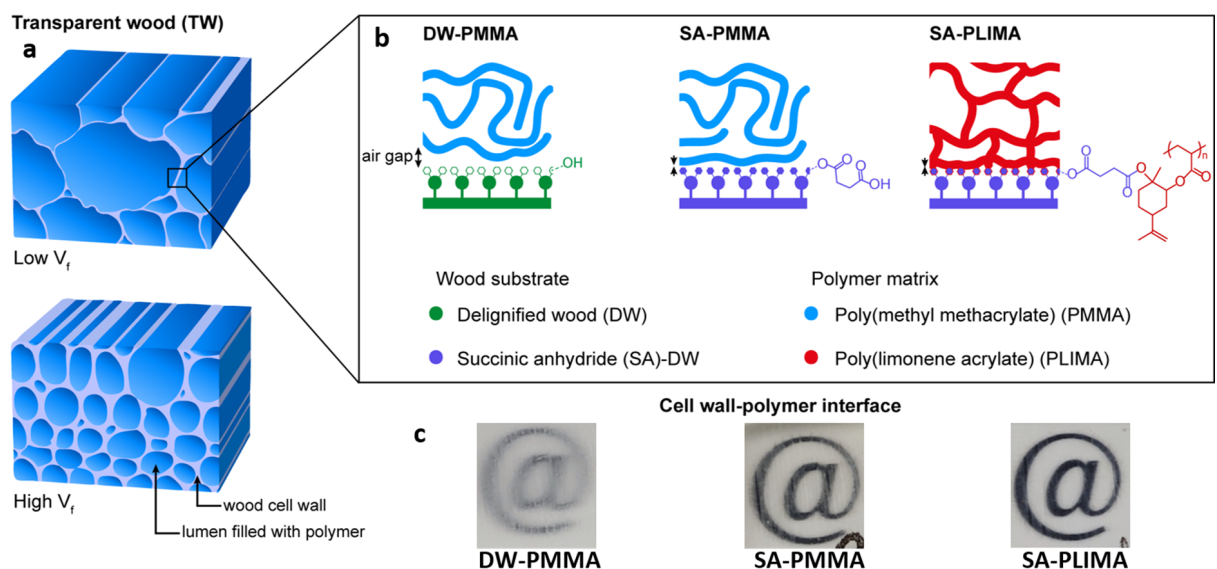

**Figure 1.** a) Illustration of transparent wood structure with various volume fraction of cellulose. b) Cell wall-polymer interfaces for three TW biocomposites termed DW-PMMA, SA-PMMA, SA-PLIMA, respectively. c) Photographs of TW samples with $V_f$ = 4.5% and thickness of 0.2 cm on top of an "@" symbol.

The variations in native wood characteristics and TW preparation methods were explored to create different materials with the aim of quantitative understanding of light-TW interaction mechanisms. For a hierarchically structured material like TW, non-uniformities of different size (*a*) both at micro- and nanoscales will contribute to the light scattering. This, will increase the optical path of the photon (of wavelength *λ*) travelling through TW, which leads to stronger absorption even for nominally delignified samples. Depending on the parameter $q = \pi a/\lambda$ we can describe elementary scattering centers in TW either by (a) ray optics ($q > 10$) or (b) Rayleigh scattering ($q < 1$). An intermediate (c) Mie regime ($1 < q < 10$) is also present, yet all phase-related effects will be averaged out by natural size/shape variations in the material. Indeed, it was numerically proven for the TW structure that solutions to Helmholtz equation yield the same result for the spatial distribution of transmitted light as geometrical optics.[40]

We consider photon propagation in the plane perpendicular to the fiber direction, where scattering has been shown to be much stronger in TW.[5] For a typical cylinder-like lumen morphology representing the polymer with RI = $n_2$ , forming an interface with a material with RI = $n_1$ (representing wood cell wall, or air gap), as shown in **Figure 2**a, the angular deviation *δ* from the initial propagation direction of normally incident photons is (from Snell's law, see Supplementary S2.1 for derivations):

$$\frac{dp}{d\delta} \sim \frac{(1-\cos\delta \cdot n_1/n_2)(n_1/n_2 - \cos\delta)^2}{(1 - 2\cos\delta \cdot n_1/n_2 + (n_1/n_2)^2)^{3/2}} \tag{1}$$

This distribution is shown in Figure 2a for PLIMA-filled lumens in case of interfaces with air or cell wall. The scattering is highly forward oriented for the latter (red curve). In this case photons exercise walk with snake-like trajectories, where the optical path is very close to the sample thickness. The backscattering probability for such an interface can be estimated from the Fresnel reflection law and is negligible ($10^{-5}$). However, for samples with microscopic air gaps, even a single interface introduces broad forward scattering (blue curve), and a relatively strong backscattering: ~ 3% for a single such interface.

At nanoscale, nano-voids and the typical distance between cellulose nanofibrils inside the cell wall is much smaller than the wavelength,[41] which leads to Rayleigh scattering with the well-known nearly isotropic $\cos^2 \delta$ angular distribution[42] (Figure 2b). This mechanism corresponds to photon random walk in TW. The co-existence of different scattering scenarios in TW samples is evidenced by time-resolved measurements of an ultra-short (150 fs) laser pulse transmission (Figure 2b, details in Figure S2a-b). Here one can distinguish early photons (with a short optical path) from the delayed photons, originating from multiple scattering. Figure 2b shows the normalized temporal profile of the measured reference pulse (red) and the scattered light pulse passed through TW ($\lambda$ = 790 nm, ballistic photons blocked). The measured TW pulse profile exhibits a fast-rising edge with duration determined by the system response and a slow decaying tail with increased full width half maxima (FWHM), in contrast to the reference signal; this is strong evidence of multiple light scattering taking place in the sample. So the narrow angle forward scattering due to RI mismatch is clearly not the only mechanism responsible for light-TW interaction.

Depending on details of the native material structure and TW preparation conditions, the amount of all these scattering centers will vary from sample to sample. Experimentally measurable quantities of the photon budget, are illustrated in Figure 2c. Here, ballistic transmittance (*BT*) corresponds to photons that were neither scattered nor absorbed. Specular reflectance (*SR*) represents photons which do not interact with the sample due to front interface reflection. Its value was determined to be $SR \approx 4\%$ (Figure S3b) for all samples used in this study, corresponding to a flat polymer-air interface. Indeed, the surface roughness of the TW biocomposite can influence backscattering,[43–45] therefore it was characterized by atomic force microscopy (AFM), Figure S1b. AFM images show that the surface roughness of the TW surface is less than 30 nm. This is much smaller than the laser wavelength (550 nm), confirming that roughness does not contribute to diffused backscattering. In the next section it is shown how photon budget quantities can be related to the elementary mechanisms in Figure 2a-b, making it possible to separate different contributions to the light-TW interaction.

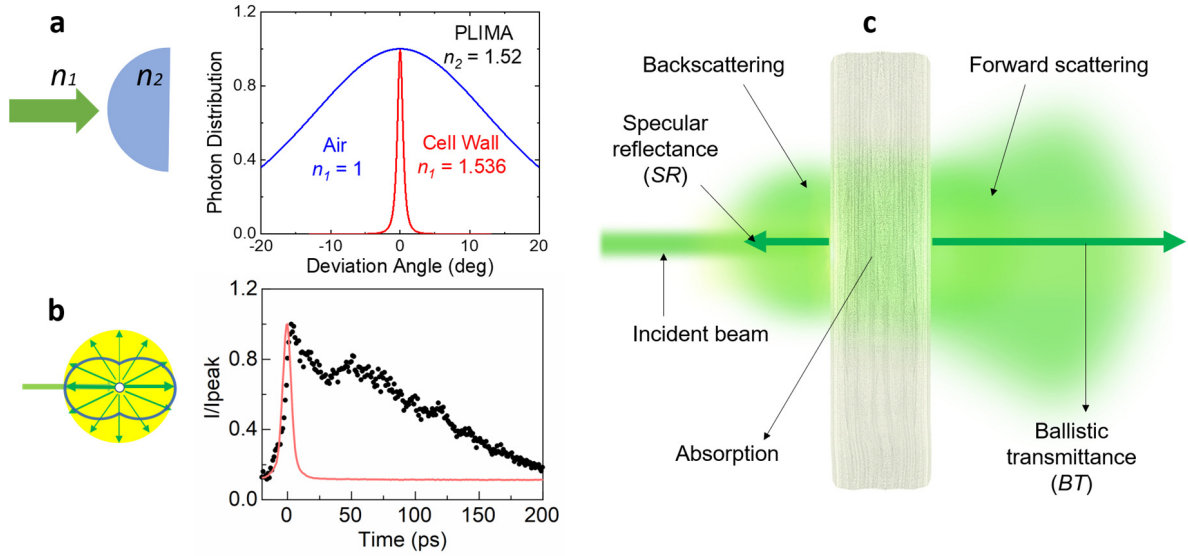

**Figure 2.** Different elementary scattering events that a photon can undergo during transport in transparent wood. a) Scattering at the polymer-filled lumen (blue, PLIMA, $n_2$ = 1.52) interface with a cell wall ($n_1$ = 1.536) or an air gap ($n_1$ = 1) and its normalized forward angular distributions. b) Rayleigh (nearly isotropic) scattering at nano-voids or nano-inclusions in the cell wall. Data for the temporal profile of a transmitted laser pulse through a 5 mm thick DW-PMMA sample, revealing a delayed part. The solid red line is the normalized input pulse. c) Measurable photon budget components.

## 2.2. Ballistic Transmittance and Wood Structure

First, *BT* was measured with a very small numerical aperture detector (Figure S3a). Results for DW-PMMA with $V_f$ =11.1% and SA-PLIMA with $V_f$ = 4.5% are shown in Figure 3a (other samples in Figure S4). To avoid sample-to-sample variations, the same specimen was gradually polished in each case. TW samples shown in **Figure 3**a represent the highest and the lowest *BT* values from this study. It is apparent that the scattering and absorption ability of all types of TW is much stronger than that for optical glass (black dots in Figure 3a).

In Figure 3a, substantial differences in how *BT* changes with increased thickness are shown for the two materials. It is apparent that structural details of the native wood sample as well as details of preparation (optical defects) dramatically affect TW scattering and absorption characteristics, where *BT* values vary in a broad range. Another important observation is that despite the quasi-random structure of TW its thickness-dependent response can be described with a constant rate (mono-exponential solid line fits). We define it as an extinction coefficient $\mu$ [cm$^{-1}$], being a sum of all scattering and absorption coefficients. In Figure 3b we show how this quantity depends on $V_f$ for two types of TW composites (raw data in Figure S4). SA-PMMA data are in-between these two extremes with a similar slope (Figure S4d).

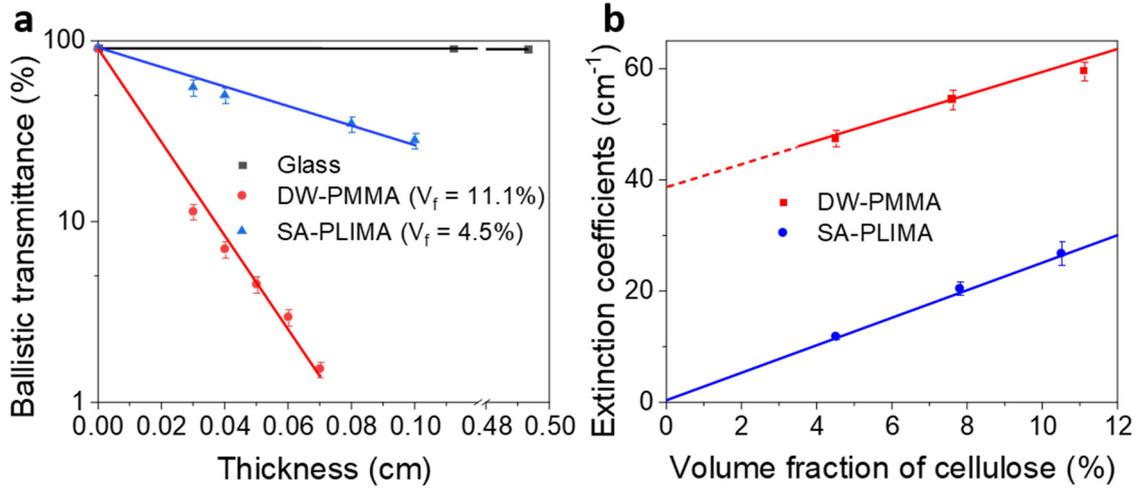

**Figure 3.** a) Ballistic transmittance (in log-linear scale) of glass, DW-PMMA ($V_f$ = 11.1%) and SA-PLIMA ($V_f$ = 4.5%) (the lowest and the highest *BT* of TW samples). The solid line is a mono-exponential fit. b) Extinction coefficient as a function of $V_f$ for these two TWs. The solid line is a linear fit.

By extrapolating linear dependencies in Figure 3b one can note that the neat SA-PLIMA sample is expected to show negligible scattering and absorption, characteristic for a pristine polymer ($V_f$ = 0%). This result in Figure 3b is because typical extinction coefficient values for amorphous polymers are similar to glass (~10$^{-2}$ cm$^{-1}$). In contrast, PMMA-based TW composites show an unexpected, large offset (dotted red line) at $V_f$ = 0%. Since PMMA is also an amorphous polymer with low extinction coefficient, it is apparent that optical defects in the form of debonding air gaps dominate light-TW interaction in this case. In **Figure 4**a, 4b, S1c we show high resolution Scanning Electron Microscope (SEM) images of typical interfaces for PMMA- and PLIMA-based TW biocomposites. This experimentally illustrates effects from differences in wood-polymer chemical interaction, as was schematically introduced in Figure 1b. It indeed appears that covalent attachment between the wood cell wall and the PLIMA polymer results in negligible amount of air gaps in SA-PLIMA TW biocomposites, see Figure 4b. Hence, the extrapolated extinction coefficient is close to 0 for $V_f$ = 0%. We therefore focus on SA-PLIMA TW biocomposites since large debonding air gaps are absent.

To better understand linear dependence in Figure 3b we performed structural characterization of TW with different $V_f$ using low magnification SEM (Figure 4c-e). The average cell wall thickness for these three samples is shown in Figure 4f, obtained from ~ 100 cells. The value for $V_f$ = 4.5% is 0.8 $\mu$m, while for samples with $V_f$ = 7.6% and 11.1% are 1.5 $\mu$m and 2 $\mu$m, i.e. linearly scaling with $V_f$. The same approach was used to measure average lumen diameters of the central pore space. It is around 38 $\mu$m for TW with $V_f$ = 4.5%, while it is 25 $\mu$m and 18 $\mu$m for $V_f$ = 7.6% and 11.1%, which is also scales approximately linearly with $V_f$. So, both the linear density of wood-polymer interfaces and of the cell wall material are proportional to $V_f$ (Figure 4f). This makes it difficult to distinguish between forward scattering from small wood-polymer RI mismatch at interfaces and random scattering contributions from nanoporosity in the cell wall. Additional analysis is therefore required.

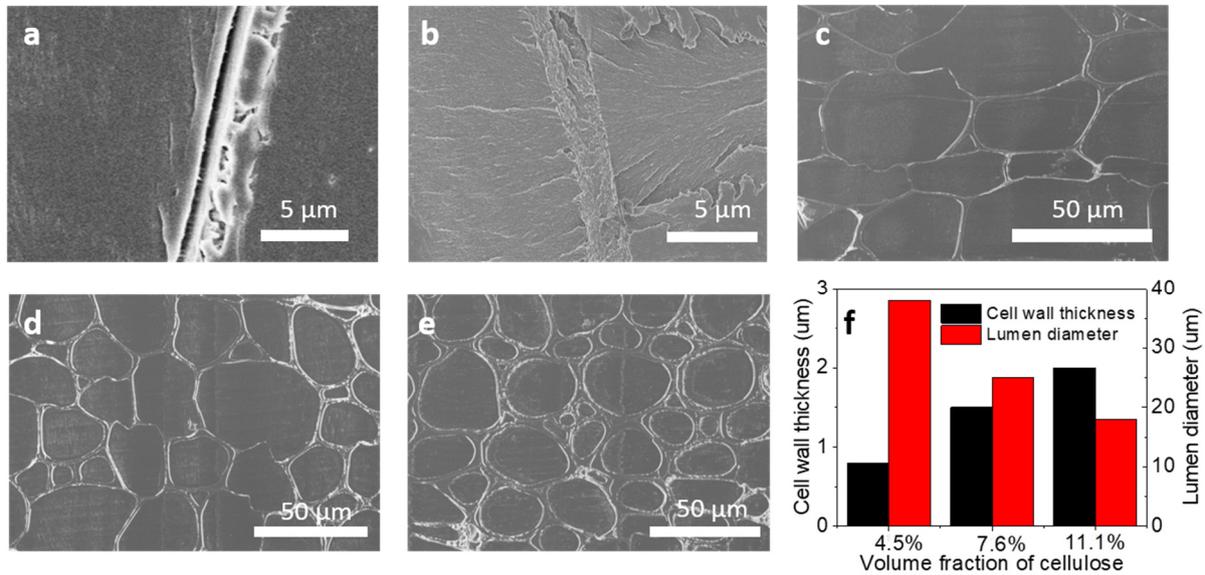

**Figure 4.** Interfaces between cell wall and polymer of a) DW-PMMA and b) SA-PLIMA TW composites at higher SEM magnification. Microstructure of DW-PMMA with c) $V_f$ = 4.5%, d) $V_f$ = 7.6% and e) $V_f$ = 11.1%, respectively, as recorded by low magnification SEM. f) comparison of cell wall thickness and lumen size for different $V_f$ of DW-PMMA TW composites.

## 2.3. Scattering and Absorption Coefficients

We developed an analytical model, which relates elementary constants of light-TW interaction to measurable photon budget quantities (cf. Figure 2c). First, in the absence of debonding gaps, $BT$ can be expressed as:

$$BT = (1 - SR)\exp(-(\mu_{sr} + \mu_{sf} + \mu_a)d) \qquad (2)$$

for a sample with thickness $d$ [cm], where $SR$ is specular reflectance, $\mu_{sr}$ is the Rayleigh (random angle) scattering coefficient, $\mu_{sf}$ is the forward scattering coefficient, $\mu_a$ is the absorption coefficient of the TW material, all in [cm$^{-1}$] units. The sum of the three is the extinction coefficient $\mu$ for SA-PLIMA.

Next, we note that the forward scattered photons in a narrow angle ($\mu_{sf}$) do not introduce sizable dispersions in the optical path. Then, another measurable quantity from the photon budget, total transmittance ($TT$), can be expressed as (see derivations in Supplementary, S2.6):

$$TT = DT_r + (1 - SR)\exp(-(\mu_{sr} + \mu_a)d) \qquad (3)$$

where $DT_r$ is the diffused transmittance for Rayleigh scattered photons. Finally, backscattering for small RI mismatch interfaces was estimated to be negligible ~ 10$^{-5}$ at a single interface. Even hundreds of such interfaces, corresponding to several mm thick TW samples used here, will not accumulate to a measurable signal (< 1%). Therefore, we can assume that the diffusely reflected light $DR$ originates solely from the Rayleigh scattering:

$$DR \approx DR_r \qquad (4)$$

Thus, there are three Equations (2)-(4) with three unknowns ($\mu_{sr}$, $\mu_{sf}$, $\mu_a$) as long as $DT_r$ and $DR_r$ can be expressed through elementary constants. To do this, we consider that after the first Rayleigh (random) scattering event, the photon trajectory can be treated as a random walk in an absorbing medium. Again, forward scattering in a narrow angle ($\mu_{sf}$) does not affect trajectories. Then a diffusion-decay equation describes propagation of such photons:

$$\frac{\partial}{\partial t}p(x,y,t) = D\left(\frac{\partial^2}{\partial x^2} + \frac{\partial^2}{\partial y^2}\right)p(x,y,t) - \beta p(x,y,t) \qquad (5)$$

where $p(x,y,t)$ is a spatial probability density to find a photon at a point with coordinates ($x$, $y$) at a time $t$, $\beta$ is the absorption rate $\beta = \mu_a c/n$, and $D$ is a photon diffusion coefficient $D = c/[3n(\mu_{sr}+\mu_a)]$, $c$ is the speed of light; $n$ is the wood refractive index. The attenuation coefficient is $\alpha = \sqrt{\beta/D} = \sqrt{3\mu_a(\mu_{sr} + \mu_a)}$. Equation (5) can be solved analytically for a point source inside a finite slab, yielding expressions for $DR_r$ and $DT_r$ (see Supplementary S2.3-2.4). In a real situation of an impinging laser beam, the source of such diffused photons is not a point, but a probabilistically spread distribution. The probability density of Rayleigh scattering for incoming photons at a point $y$ is (1-$SR$)$\mu_{sr}$exp[-($\mu_{sr}$+$\mu_a$)$y$] for photons not absorbed in the material before $y$ and not specularly reflected. Convoluting such a distributed source with analytical solutions of Equation (5) for a point source, we obtain the final expressions (Supplementary S2.5)

$$DT_r = \frac{\mu_{sr}(1-SR)((\mu_{sr}+\mu_a+\alpha)e^{(\alpha-\mu_{sr}-\mu_a)d}-(\mu_{sr}+\mu_a-\alpha)e^{-(\alpha+\mu_{sr}+\mu_a)d}-2\alpha)}{2\sinh(\alpha d)(\alpha^2-(\mu_{sr}+\mu_a)^2)} \qquad (6)$$

$$DR_r = \frac{\mu_{sr}(1-SR)\left((\mu_{sr}+\mu_a+\alpha)e^{-\alpha d}-(\mu_{sr}+\mu_a-\alpha)e^{\alpha d}-2\alpha e^{-(\mu_{sr}+\mu_a)d}\right)}{2\sinh(\alpha d)(\alpha^2-(\mu_{sr}+\mu_a)^2)} \qquad (7)$$

These are general expressions relating measurable photon budget quantities to elementary scattering and absorption constants in a material. This should allow us to quantify the relative contributions from mechanisms in Figure 2a-b.

## 2.4. Photon Budget of TW

To apply this theory to TW, spectrally-resolved *TT* and total reflectance *TR,* which is a sum of *SR* and *DR*, were measured in the set-up shown in Figure S3c and S3d. The absorption was obtained as:

$$A = 1 - TT - TR \qquad (8)$$

An example of photon budget for a SA-PLIMA sample is shown in **Figure 5**a (for other samples in Figure S5). It is seen that in the visible spectral range there is almost no wavelength-dependence for any of the quantities. *TT* and *TR* results at wavelength of 550 nm are summarized in Figure 5b for two different samples (other samples in Figure S6). From Figure 5b one can see that for each type of TW sample, both *TT* and *TR* in a log-scale are roughly proportional to the sample thickness (increasing for *TR*, and decreasing for *TT*) in agreement with our previous result for *TT*.[46] Again, one can note that extrapolation of the *TR* dependence to zero thickness approximately yields *SR* values (4%) for the SA-PLIMA case, whereas they are substantially higher for DW-PMMA sample, attributed to scattering from internal air gaps.

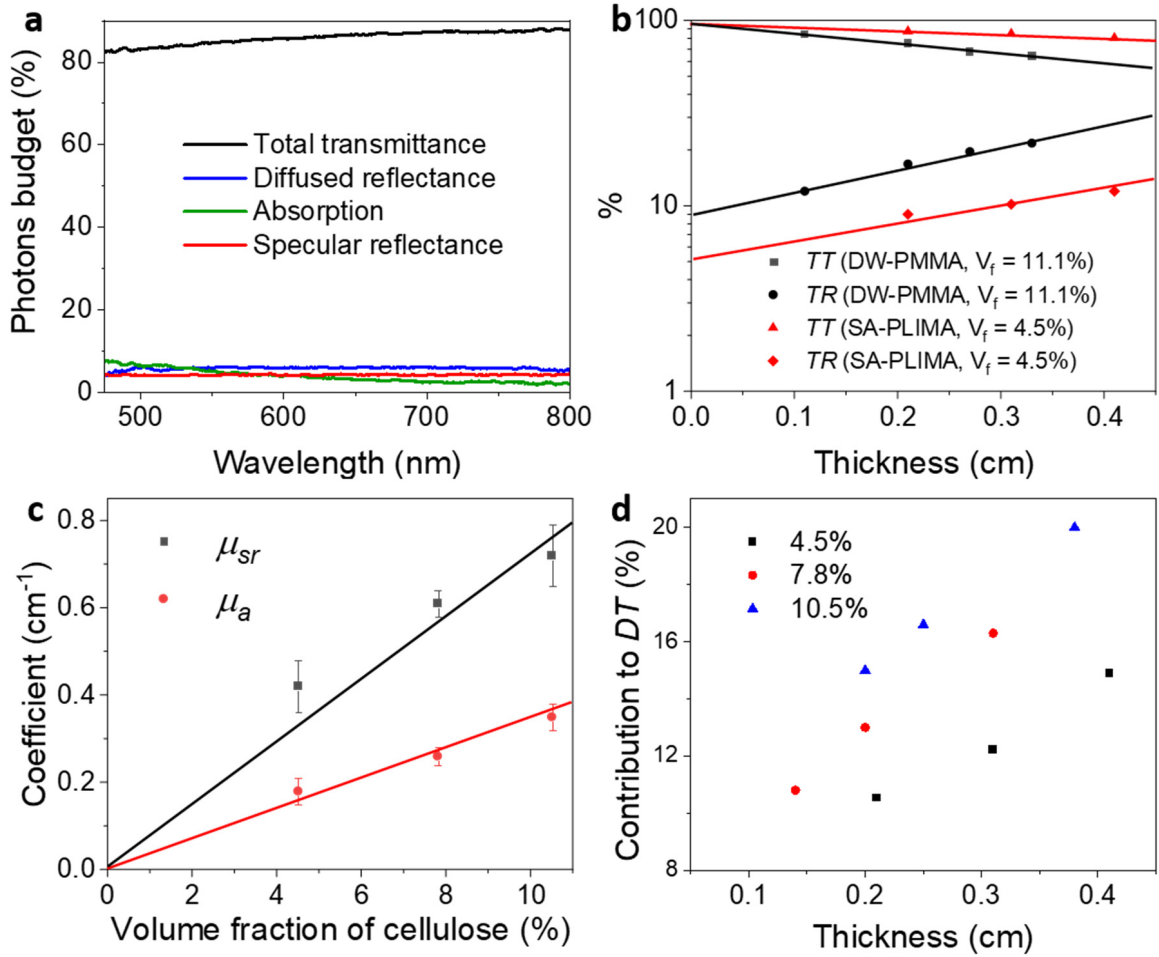

**Figure 5.** a) Spectrally resolved photon budget for SA-PLIMA sample with $V_f$ = 4.5% and thickness = 0.31 cm. b) total transmittance *TT* and total reflectance *TR* (sum of specular and diffused reflectance) of DW-PMMA ($V_f$ = 11.1%) and SA-PLIMA ($V_f$ = 4.5%) samples with various thicknesses. c) Rayleigh (random) scattering and absorption coefficients for SA-PLIMA TW. Note that values are low compared with the extinction coefficient $\mu$. d) Contribution of diffused transmittance from Rayleigh (random) scattering ($DT_r$) to the total diffused transmittance (*DT*) for different cellulose volume fraction samples.

Substituting experimentally measured photon budget quantities into Equations (2)-(4), and, numerically solving this system of equations, $\mu_{sr}$ and $\mu_a$ of the SA-PLIMA TW were obtained. Results, as averaged values for all sample thicknesses, are presented in Figure 5c. They reveal that both coefficients (random Rayleigh scattering and absorption) scale linearly with $V_f$, while absolute values remain low. For example, $\mu_{sr} = 0.46 \pm 0.06$ cm$^{-1}$ and $\mu_a = 0.18 \pm 0.03$ cm$^{-1}$ for $V_f$ = 4.5%, which are only small fractions of $\mu = 11.9 \pm 0.5$ cm$^{-1}$. This means that the Rayleigh (random) scattering process from cell wall nanopores and inter-fibril voids plays less important role in light-TW interaction than forward scattering from polymer/wood RI mismatch.

The origin of the minor absorption in TW samples can be assigned to residual lignin content. Its value was estimated to be low for the delignified template (≈ 2 wt. % vs. 24.9% lignin for native wood[24]). To verify this, we prepared non-delignified TW samples and measured the photon budget (Figure S5d). One can see that absorption becomes much more pronounced and it reveals a spectral dependence. Stronger absorption in the blue range is a typical signature of lignin absorption.[47] Thus, we can conclude that the forward-scattering caused by the polymer-wood RI mismatch dominates total optical response of delignified TW samples.

So, to obtain TW with optical characteristics close to pristine polymer or glass, such as low haze, the problem of wood-polymer RI mismatch should be addressed first. It can be accomplished by adding cellulose nanofibrils to a monomer, or by utilizing a two-component polymer, such as OSTE,[48] with a tunable stoichiometric ratio to fine tune RI to that of the cell wall.

The nanoporosity or heterogeneity of the cell wall, however, cannot be completely ignored. Although the absolute value of the Rayleigh (random) scattering coefficient appears to be 20-30 times lower than $\mu_{sf}$, it is still an order of magnitude higher than for a pristine polymer. More importantly, this random angle scattering contributes to haze already after a single event, while multiple events are required for the forward scattering to achieve beam deviation >2.5°, as per haze definition.[49] This fact is illustrated in Figure 5d, where the contribution from Rayleigh scattering to the total diffused transmittance ($DT$) is shown as $DT_r/DT$. Clearly, it reaches non-negligible 20% for the thickest sample (0.38 cm) with the highest cellulose content (10.5%). Therefore, reduction of small-scale heterogeneities in the wood cell wall (nanopores, inter-fibril voids etc.) will further reduce haze towards an optically clear wood.

## 3. Conclusions

The main factor influencing light scattering in transparent wood (TW) is forward scattering caused by refractive index (RI) mismatch between the cell wall and the polymer. In this investigation, we also explain in detail how light-TW interaction is influenced by sample thickness and volume fraction of cellulose, linear density of interfaces, interface compatibility (wood-polymer interface debond gaps), lignin content, and the RI of the polymer matrix. Photon budget measurements were combined with photon diffusion theory for repeated scattering events to estimate extinction coefficients, Rayleigh (random) scattering, forward scattering, and absorption coefficients of TW. The results show that the lignin residue causes only minor absorption. Rayleigh (random) scattering is low and mainly due to small scale defects in the wood cell wall phase in TW. Light scattering in TW is indeed dominated by forward scattering from RI mismatch between the cell wall and the polymer. If air voids, i.e. debonding air gaps between cell wall and polymer, are not minimized at the fabrication stage they will completely dominate scattering in TW.

These quantitative results for TW are essential for understanding the light-TW interaction and provide a guideline to future TW fabrications strategies. For example, decreasing the lignin content in order to lower the light absorption, modifying the wood substrate in order to improve the compatibility between cell wall and polymer, finding a proper polymer with suitable RI to improve the transmittance, etc. The method applied in this study could also be used for extracting scattering and absorption coefficients for other types of turbid composites from photon budget measurements.

## 4. Experimental Section

*Fabrication of TW:* In the present work, TW with various thicknesses made from various densities of balsa wood were fabricated on the basis of our previous methods[24] with minor modifications. Balsa wood (oven dried density: 79, 137 and 183 kg m$^{-3}$, purchased from Wentzels Co. Ltd., Sweden) samples were cut with dimension of 1.5 × 1.5 cm from the plane cut veneer (with the ray cells in the thickness direction) with various thicknesses (from around 0.1 to 0.7 cm). Firstly, samples were delignified at 80 °C using 1 wt % of sodium chlorite (NaClO$_2$, Sigma-Aldrich) in an acetate buffer solution (pH 4.6) for 6 h (12 h for the sample thickness above 0.3 cm, to make sure there are similar composition in the wood substrates). Then the delignified wood (DW) samples were washed with deionized water, ethanol, and finally acetone, each step repeated for 5 times. DW substrates were immersed in succinic anhydride (SA, ≥99%) under 130 °C for 30 minutes following our previous method.[37] After the reaction, the samples were washed in acetone for 5 times under vacuum. In order to remove the color from the reaction, the substrates were repeated the delignification process for 30 minutes. Methyl methacrylate (MMA) pre-polymerization was performed in a round bottom flask at 75 °C for 15 min with 2,2′-Azobis(2-methylpropionitrile) (AIBN, 0.3 wt %, Sigma-Aldrich) as initiator and terminated with ice-water bath. The poly(limonene acrylate) (PLIMA) was prepared according to our previous method.[37] The DW or SA treated DW substrates were infiltrated with pre-polymerized MMA or LIMA monomer under vacuum for 4 h (12 h for sample thickness above 0.3 cm). Finally, the samples were sandwiched between 2 glass slides sealed in aluminum foil and polymerized in oven, with the following heating steps: start at 35 °C, with 5 °C interval until 70 °C, each step is kept for 4 h to gradually complete the polymerization process while minimizing polymer shrinkage. The obtained TW samples are named as DW-PMMA (DW + PMMA), SA-PMMA (SA treated DW + PMMA), SA-PLIMA (SA treated DW + PLIMA), non-delignified TW (native wood + PMMA).

*Characterization:* Volume fraction of holocellulose ($V_f$, cellulose and hemicelluloses, for simplicity, cellulose used in this work) was calculated as

$$V_f = \frac{W_f \times \rho_c}{\rho_f}$$

where $V_f$ is the volume fraction of cellulose, $\rho_c$ is the density of the composite, $\rho_f$ is the density of cellulose (1500 kg m$^{-3}$), $W_f$ is the weight fraction of cellulose. Atomic Force Microscopy (AFM) topographic height images of the TW surface were recorded in air using a ScanAsyst mode on a MultiMode 8 atomic force microscope system (Bruker, Santa Barbara, CA, USA) with a scan area of 5 $\mu$m × 5 $\mu$m. The cross sections of the samples were observed with a Field-Emission Scanning Electron Microscope (SEM, Hitachi S-4800, Japan). Ultra-fast time resolved measurement was performed using the setup as shown in Figure S2a. The setup was verified with colloidal suspension of silica particles of different diameters, which time-resolved signal is shown in Figure S2b. Longer tail can be observed in the case of smaller nanoparticles signifying multiple scattering (Rayleigh regime). Ballistic transmittance (*BT*) was measured with a "bucket" detector[50] as shown in Figure S3a, where only the ballistic transmitted photons could be collected by the detector, and the scattered part was blocked by the black bucket. Specular reflectance (*SR*) was measured using the setup shown in Figure S3b, where the sample was set in the middle of a round plate, the input angle of the beam was 6º and the detector was set in the corresponding position on the edge of the round plate (radius = 13.6 cm). The aperture of the detector was similar to the directly reflected beam size. Angle-integrated total transmittance (*TT*) and diffused transmittance (*DT,* with scattering angle >2.5˚) of the samples was measured with an integrating sphere according to ASTM D1003 "Standard Test Method for Haze and Luminous Transmittance of Transparent Plastics" as shown in Figure S3c and S3e. Total reflectance (*TR*) was obtained by placing the sample on the back opening of the integrating sphere as shown in Figure S3d. In this case, all the reflected photons from both surface and the interior of the TW composite could be collected by the detector through the integrating sphere.

**Supporting Information**

Supporting Information is available.

**Acknowledgements**

We acknowledge funding from KTH Royal Institute of Technology and European Research Council Advanced Grant (No. 742733) Wood NanoTech, and funding from Knut and Alice Wallenberg foundation through the Wallenberg Wood Science Center at KTH Royal Institute of Technology.

**Conflict of Interest**

The authors declare no conflict of interests.

Received:

Revised:

Published online:

Supplementary Information for

# Photon Walk in Transparent Wood: Scattering and Absorption in Hierarchically Structured Materials

*Hui Chen[1], Céline Montanari[1], Ravi Shanker[1], Saulius Marcinkevicius[2], Lars A. Berglund,[1]*

*Ilya Sychugov\*[2]*

1, Wallenberg Wood Science Center, Department of Fiber and Polymer Technology, KTH Royal Institute of Technology, Teknikringen 56, 100 44 Stockholm, Sweden

2, Department of Applied Physics, School of Engineering Sciences, KTH Royal Institute of Technology, Hannes Alfvens väg 12, 114 19 Stockholm, Sweden

\* E-mail: ilyas@kth.se

**CONTENTS**

# Section S1. Experimental details

**Refractive index (RI)** of succinic anhydride (SA) treated delignified wood (DW) substrate was measured using immersion liquid method combined with transmission model as we developed in our previous research,[S1] the result is shown in Fig. S1a, where the RI of SA-DW was obtained as 1.533 at wavelength of 589 nm.

**Atomic Force Microscopy (AFM)** was applied to characterize the roughness of the TW sample surface with a topographic height images as shown in Fig. S1b, where it shows that the roughness of the sample is around 30 nm, which is much smaller compared with the laser wavelength (550 nm) used in this study.

**Field-Emission Scanning Electron Microscope (FE-SEM)** was used for charactering the interface between cell wall and polymer for SA-PMMA TW samples as shown in Fig. S1c, where a number of air voids were appeared at the interface.

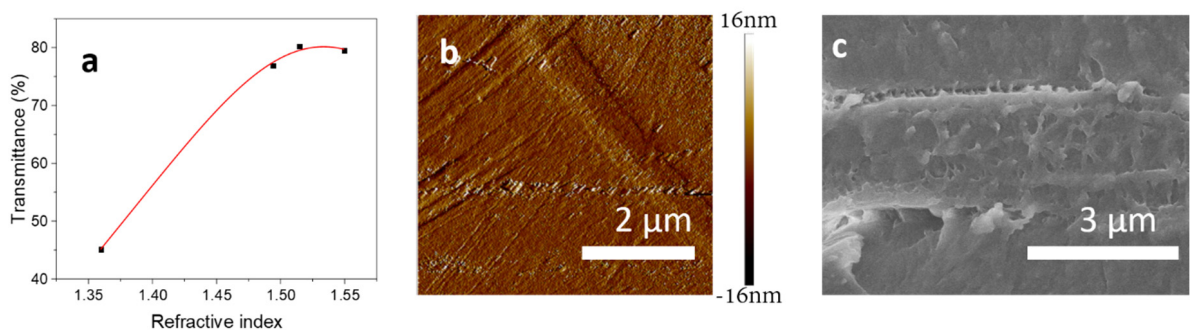

**Fig. S1.** a) Transmission model fit of the transmittance values at wavelength 550 nm for succinic anhydride treated delignified wood substrate. b) AFM image of the TW surface. c) Interface between cell wall and polymer of SA-PMMA TW sample.

**Ultra-fast time resolved measurement** was performed using the setup as shown in Fig. S2a. This was carried out with 150 fs pulses from a Ti:sapphire laser ( $\lambda$= 790 nm) with a synchroscan streak camera as a detector (~ 4 ps resolution). Ballistic photons were blocked using a beam blocker. Colloidal suspension of silica particles of different diameters was measured using this setup as a reference. The result is shown in Fig. S2b. Smaller particles, which scatter more isotropically (Rayleigh) introduce a clear delay signal.

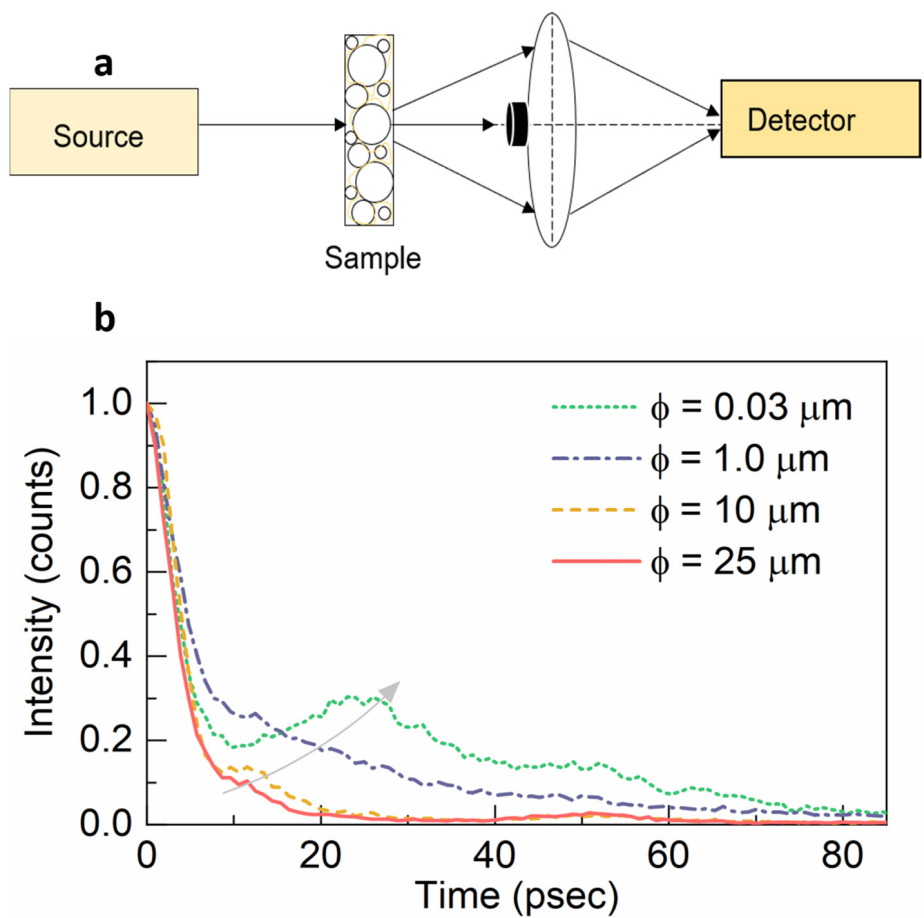

**Fig. S2.** a) Setup for ultra-fast time resolved measurement with a streak camera. b) Results for reference samples of colloidal suspension of silica particles of different diameters.

**Ballistic transmittance** at wavelength of 550 nm was performed using the setup as shown in Fig. S3a, where a "bucket" detector[S2] was applied in order to block the diffused photons from the output surface of the sample.

**Specular reflectance** was obtained using the setup as shown in Fig. S3b, where a bucket was placed in front of the detector (integrating sphere). The input beam angle was 6° to normal, with the detector in the corresponding reflection angle.

**Total transmittance and diffused transmittance** were measured using the setups as shown in Fig. S3c and S3d according to ASTM D1003 "Standard Test Method for Haze and Luminous Transmittance of Transparent Plastics",[S3] respectively, where the sample was located at the input port of the integrating sphere, respectively.

**Total reflectance** was measured using the setup as shown in Fig. S3e. The sample was located at the output port of the integrating sphere.

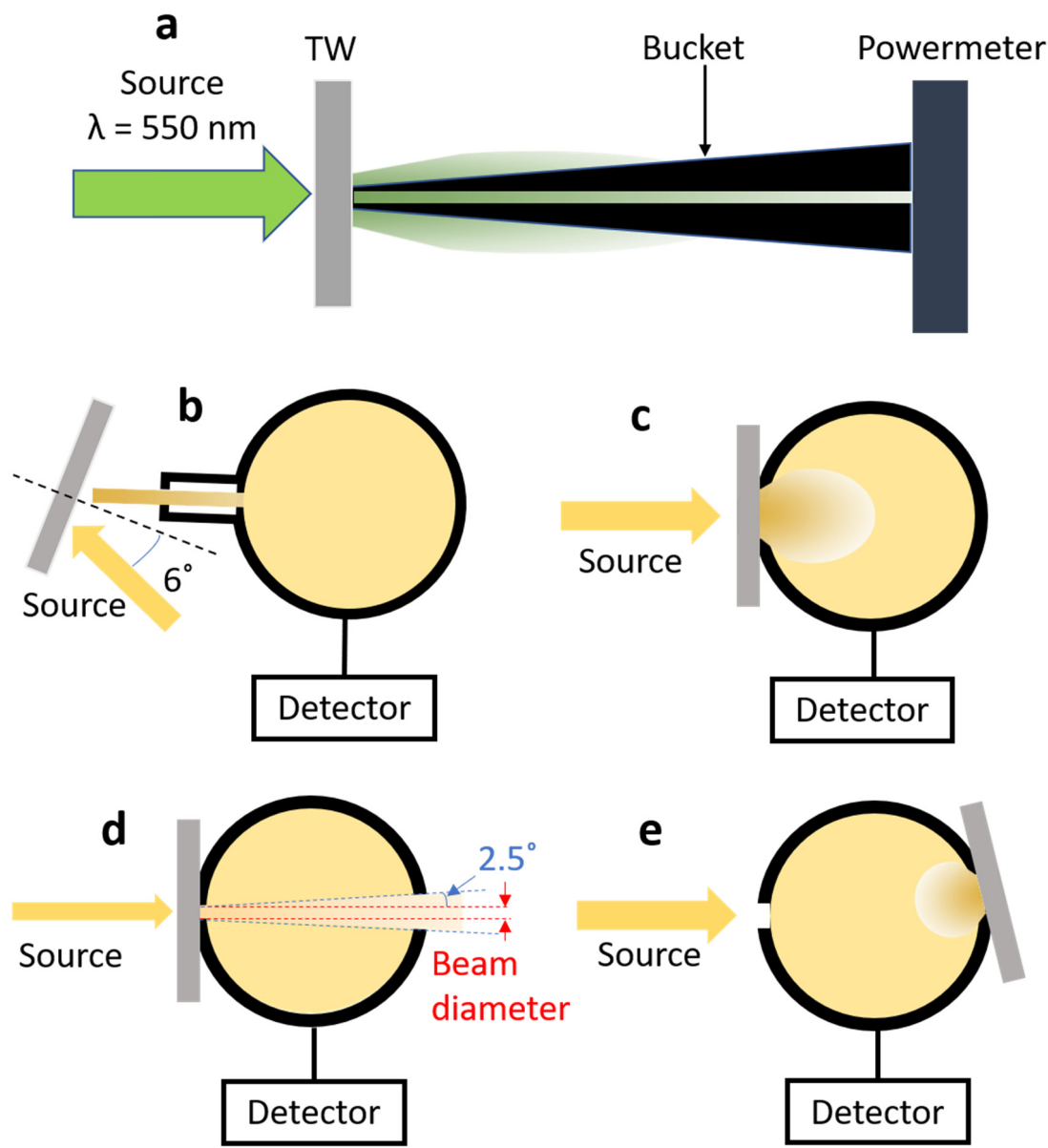

**Fig. S3.** Measurement setups for a) ballistic transmittance, b) specular reflectance, c) total transmittance d) diffused transmittance and e) total reflectance, respectively.

Ballistic transmittance of DW-PMMA, SA-PLIMA, and SA-PMMA TW samples is shown in Fig. S4a, S4b and S4c, respectively. It is clear that with increased sample thickness or volume fraction of cellulose ($V_f$), the value of *BT* is decreased for all cases, which leads to a higher extinction coefficient. One example for SA-PMMA is shown in Fig. S4d.

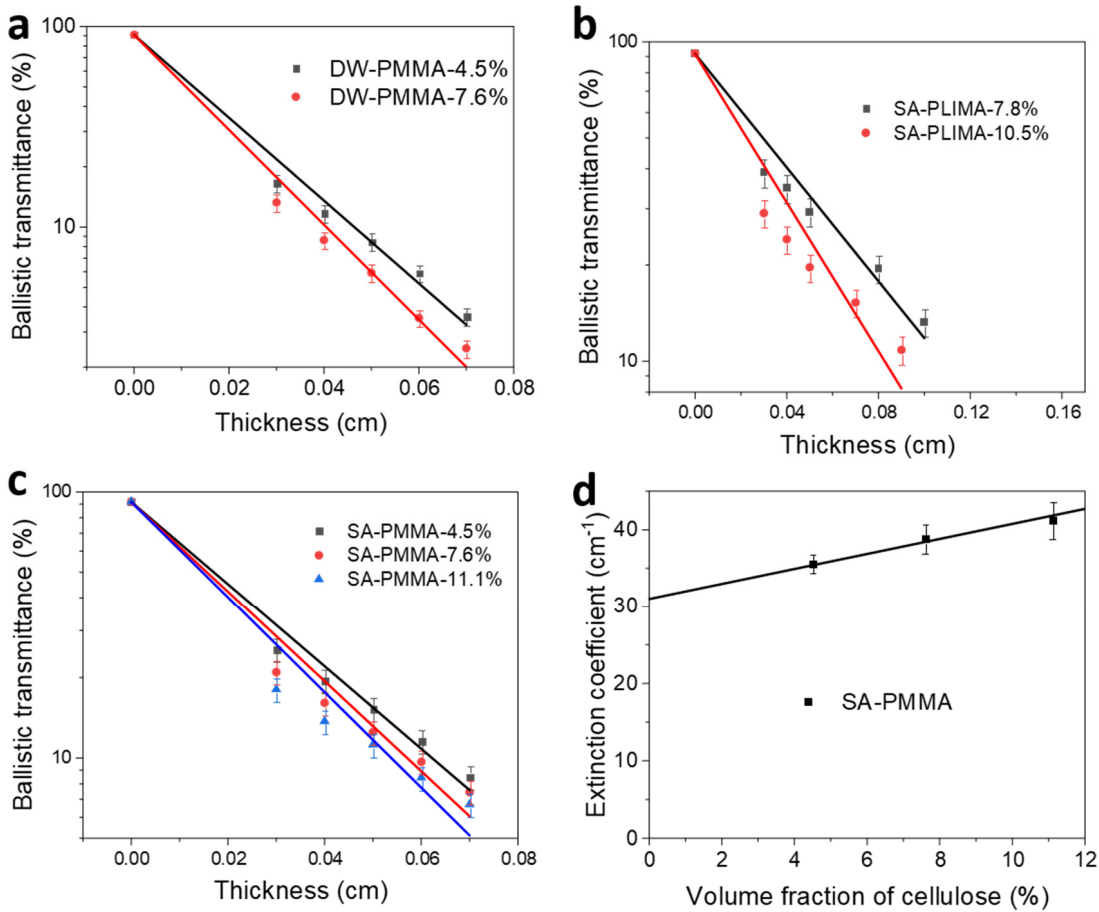

**Fig. S4.** Ballistic transmittance of a) DW-PMMA samples with $V_f$ = 4.5% and 7.6%, b) SA-PLIMA samples with $V_f$ = 7.8% and 10.5%, and c) SA-PMMA samples with $V_f$ = 4.5%, 7.6% and 11.1%, respectively. d) extinction coefficients for SA-PMMA samples with $V_f$ = 4.5%, 7.6% and 11.1%, respectively.

Spectrally resolved photon budget of various TW samples, where interface reflection (Fresnel reflection, $SR \approx 4\%$ for all the samples) is not included in the diagrams. With the same amount of $V_f$ but different sample thicknesses as shown in Fig. S5a and S5b, the light absorption and diffused reflectance (*DR*) are lower, while total transmittance (*TT*) is higher for samples with thickness of 0.17 cm compared with the sample thickness 0.27 cm. With similar thickness (around 0.27 cm), when $V_f$ increased (11.1%), the absorption and *DR* increased, while *TT* decreased as shown in Fig. S5c. Both cases are due to the higher content of wood substrate in the thicker or higher $V_f$ samples. When comparing same thickness but different lignin content (Fig. S5a and S5d), *TT* decreased while absorption and *DR* increased. This could be due to the high absorption of lignin especially in the lower wavelength range.

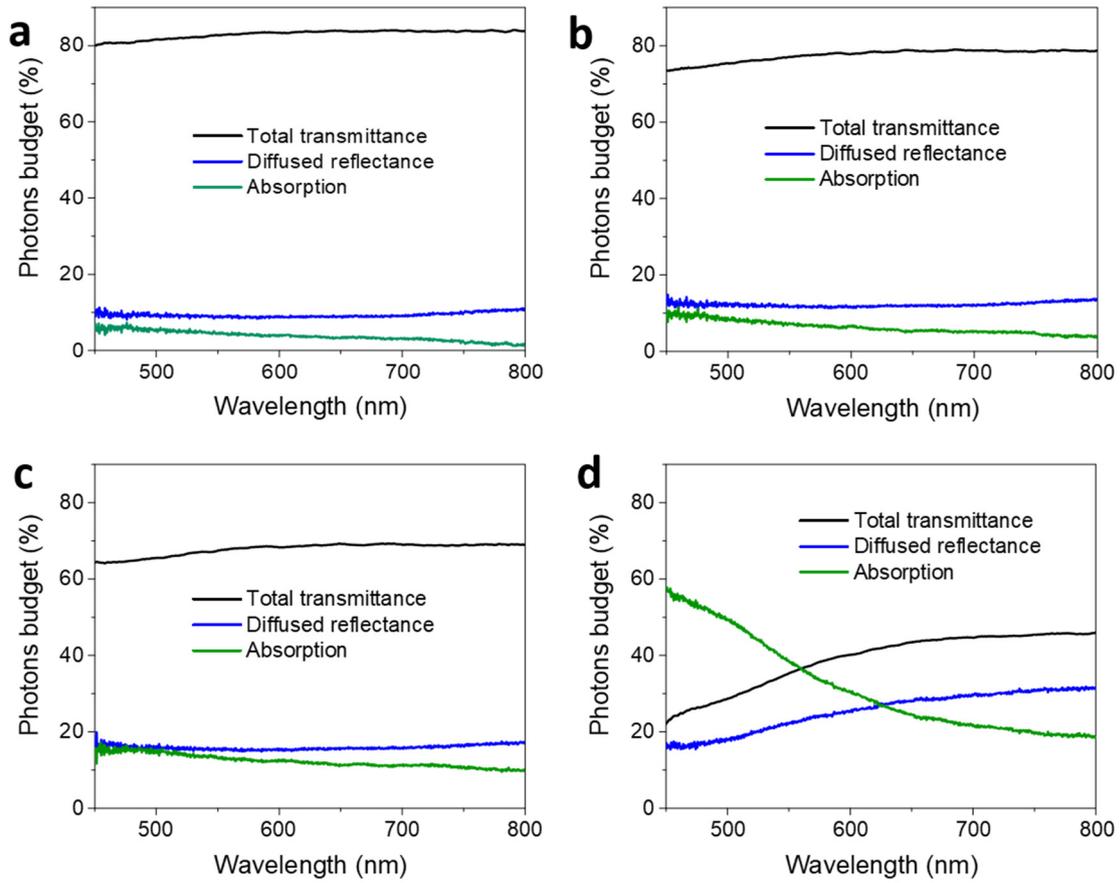

**Fig. S5.** Spectrally resolved photon budget for DW-PMMA ($V_f$ = 4.5%) with thickness of a) 0.17 cm and b) 0.27 cm, c) DW-PMMA, $V_f$ = 11.1%, 0.27 cm, and d) non-delignified TW (0.27 cm), respectively.

The results of *TT* and *TR* at wavelength of 550 nm for the TW samples are shown in Fig. S6, where 3 types of TW samples with various $V_f$ and thicknesses were showed. From the results one can see that with the sample thickness increased, *TT* is decreased while *TR* is increased (both in a log-scale). This result for *TT* is in agreement with our previous finding.[S4]

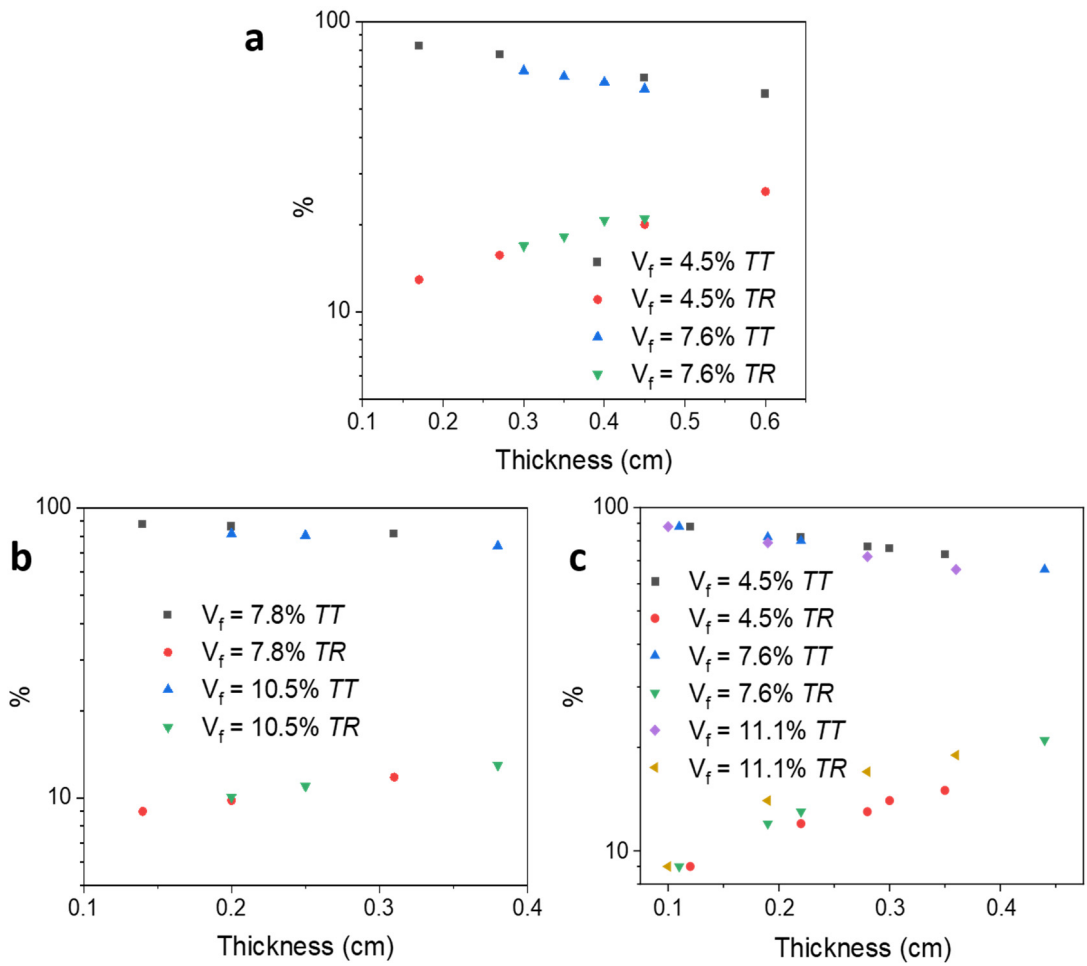

**Fig. S6.** Total transmittance and reflectance at wavelength of 550 nm for a) DW-PMMA, b) SA-PLIMA and c) SA-PMMA samples with various $V_f$ and sample thicknesses.

# Section S2. Theoretical Derivations

## S2.1 Angular distribution of photons in Fresnel refraction

For a uniform photon flux $\frac{dp}{dy} = c$ impinging on the circular interface between materials with $n_2$ and $n_1$ we are looking for distribution of the outgoing photons over the angle $\delta$, which is a deviation from the incoming angle $\theta$:

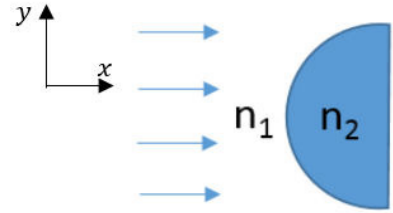

$\delta = \theta - \theta' = \theta - \arcsin(n_1 \sin(\theta)/n_2)$, from Snell's law.

One can re-write as

$$\theta = \arctan\left(\frac{\sin\delta}{\cos\delta - n_1/n_2}\right)$$

And the derivative:

$$\left|\frac{d\theta}{d\delta}\right| = \frac{1 - \cos\delta \cdot n_1/n_2}{1 - 2\cos\delta \cdot n_1/n_2 + (n_1/n_2)^2}$$

Original distribution of photons over the incoming angle $\theta$ (for a circle with a radius $R$):

$$\left|\frac{dp}{d\theta}\right| = \frac{dp}{dy} \cdot \frac{dy}{d\theta} = c \cdot R \cdot \cos\theta = c' \cdot \cos\theta = \frac{c' \cdot (n_1/n_2 - \cos\delta)^2}{\sqrt{1 - 2\cos\delta \cdot n_1/n_2 + (n_1/n_2)^2}}$$

Then the final distribution of photons over the deviation $\delta$:

$$\left|\frac{dp}{d\delta}\right| = \frac{dp}{d\theta} \cdot \frac{d\theta}{d\delta} = \frac{c' \cdot (1 - \cos\delta \cdot n_1/n_2)(n_1/n_2 - \cos\delta)^2}{(1 - 2\cos\delta \cdot n_1/n_2 + (n_1/n_2)^2)^{3/2}}$$

For refractive index values $n_2(OSTE) = 1.56$ and $n_1 = 1.52$, which corresponds to OSTE/CW interface:

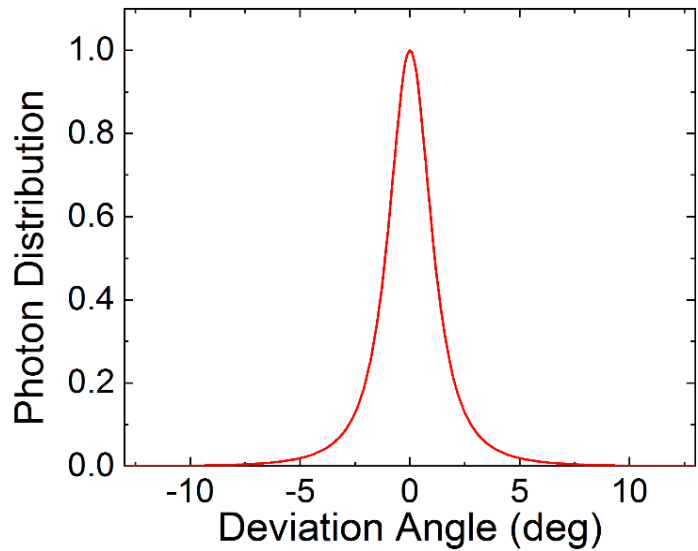

## S2.2 Diffusion model for a semi-infinite slab

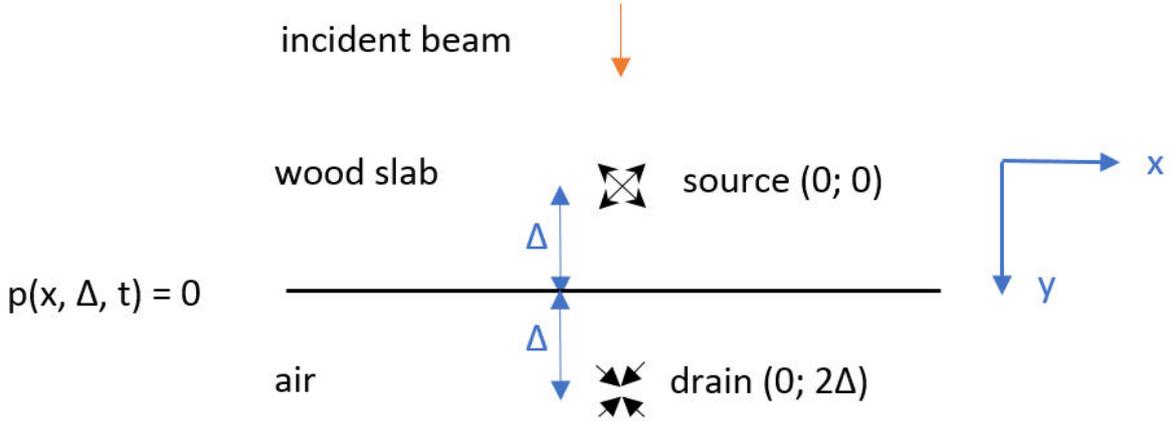

Photons enter wood slab under normal incidence, where they experience Fresnel and Rayleigh scattering and low absorption. After the first Rayleigh scattering event their trajectory can be treated as a random walk. Diffusion-decay equation describes propagation for such photons:

$$\frac{\partial}{\partial t}p(x,y,t) = D \cdot \left(\frac{\partial^2}{\partial x^2} + \frac{\partial^2}{\partial y^2}\right)p(x,y,t) - \beta \cdot p(x,y,t)$$

where $p(x,y,t)$ is a spatial probability density to find photon at a point with coordinates $(x,y)$ at a time $t$, $\beta$ is the absorption rate $\beta = \mu_a c/n$, and $D$ is a diffusion coefficient $D = c/3n(\mu_{sr} + \mu_a)$. Absorption coefficient is $\mu_a$, and random walk scattering coefficients is $\mu_{sr}$; $c$ is a speed of light; $n$ is a wood refractive index. Attenuation coefficient: $\alpha = \sqrt{\beta/D} = \sqrt{3\mu_a(\mu_a + \mu_{sr})}$.

For a point source in the semi-infinite slab at a distance $\Delta$ from an interface, where the probability density is zero $p(x, \Delta, t) = 0$ because photons leave the system at this interface, the solution is:

$$p(x,y,t) = \frac{\exp\left(-\frac{x^2+y^2}{4Dt}\right) - \exp\left(-\frac{x^2+(y-2\Delta)^2}{4Dt}\right)}{4\pi Dt} \exp(-\beta t)$$

by the method of images. By integrating this photon spatial distribution inside the slab: $y \in (-\infty; \Delta)$, and $x \in (-\infty; \infty)$ one gets probability for a photon to be inside the material at a time $t$:

$$P(t) = \exp(-\beta t) \cdot \text{erf}\left(\frac{\Delta}{\sqrt{4Dt}}\right)$$

which is a decaying function due to losses through absorption and outflow. Taking derivative gives a temporal probability density for a photon to vanish at a time $t$:

$$q(t) = -\frac{\Delta}{\sqrt{4\pi Dt^3}} \exp\left(-\beta t - \frac{\Delta^2}{4Dt}\right) - \beta \exp(-\beta t)\,\text{erf}\left(\frac{\Delta}{\sqrt{4Dt}}\right)$$

which has two terms: the first is the outflow through the interface (Fick's law):

$$|f(x,t)| = D\frac{\partial}{\partial y}p(x,y,t)|_{y=\Delta} = \frac{\Delta}{4\pi Dt^2}\exp\left(-\beta t - \frac{x^2+\Delta^2}{4Dt}\right)$$

integrated over $x$-coordinate:

$$f(t) = \frac{\Delta}{\sqrt{4\pi Dt^3}}\exp\left(-\beta t - \frac{\Delta^2}{4Dt}\right)$$

and the second is the probability density to get absorbed in the material at a time $t$. Function $|q(t)|$ is a pdf normalized to unity, as can be directly verified by time integration. This means that eventually a photon will be either absorbed or will leave the material through the interface.

Integrating $f(x,t)$ over time, instead of space, gives spatial distribution along the interface:

$$f(x) = \sqrt{\frac{\beta \Delta^2}{D\pi^2(x^2 + \Delta^2)}} K_1\left(\sqrt{\frac{\beta}{D} \cdot (x^2 + \Delta^2)}\right)$$

Total fraction of the transmitted photons can be obtained by integrating $f(x,t)$ over the time and space, which yields a simple and well-known result used previously[S4]:

$$F = \exp\left(-\Delta\sqrt{\beta/D}\right)$$

Quantities $f(t)$, $f(x)$, and $F$ can be measured by a streak-camera, a fiber placed close to the sample at different points, and by an integrating sphere, correspondingly.

### S2.3 Diffusion model for the finite slab (source in the middle)

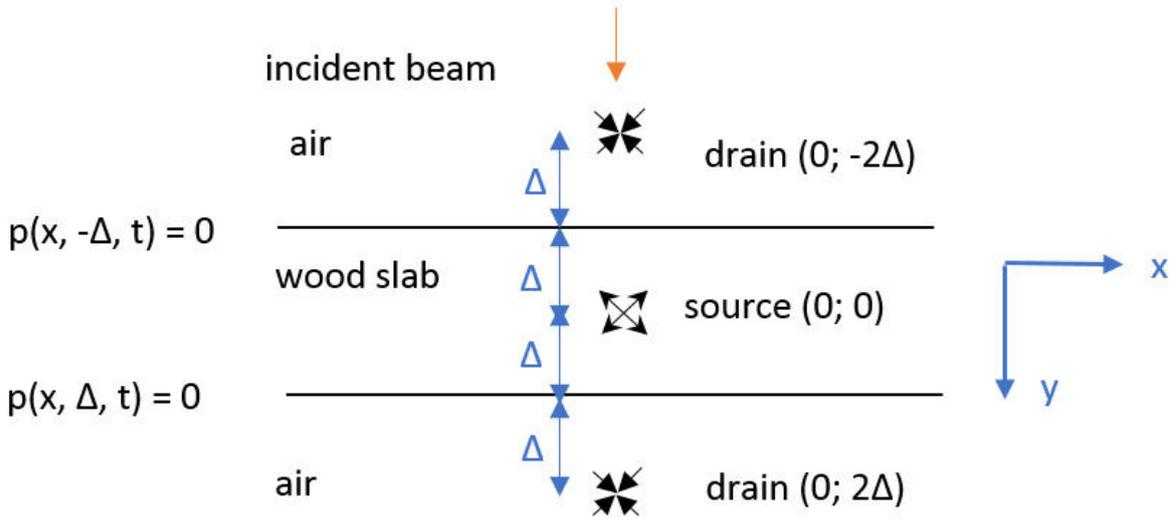

In reality the slab has a second interface. For a slab with thickness $d$ first assume the source is in the middle of the slab $d = 2\Delta$. Then one should add boundary condition $p(x, -\Delta, t) = 0$ and the solution

$$p^*(x, y, t) = \frac{\exp\left(-\frac{x^2 + y^2}{4Dt}\right) - \exp\left(-\frac{x^2 + (y - 2\Delta)^2}{4Dt}\right) - \exp\left(-\frac{x^2 + (y + 2\Delta)^2}{4Dt}\right) + \cdots}{4\pi Dt} \exp(-\beta t)$$

And one should keep adding mirror sources to infinity to enforce that boundary conditions are zero. Integrating $y \in (-\Delta; \Delta)$, and $x \in (-\infty; \infty)$ gives probability for a photon to be inside the slab at $t$

$$P^*(t) = 2\exp(-\beta t) \cdot \left(\text{erf}\left(\frac{\Delta}{\sqrt{4Dt}}\right) - \text{erf}\left(\frac{3\Delta}{\sqrt{4Dt}}\right) + \text{erf}\left(\frac{5\Delta}{\sqrt{4Dt}}\right) - \cdots\right)$$

Taking derivative:

$$q^*(t) = -\left(\frac{\Delta}{\sqrt{\pi Dt^3}} \exp\left(-\beta t - \frac{\Delta^2}{4Dt}\right) - \frac{3\Delta}{\sqrt{\pi Dt^3}} \exp\left(-\beta t - \frac{9\Delta^2}{4Dt}\right) + \frac{5\Delta}{\sqrt{\pi Dt^3}} \exp\left(-\beta t - \frac{25\Delta^2}{4Dt}\right) - \cdots\right)$$
$$- 2\beta e^{-\beta t}\left(\text{erf}\left(\frac{\Delta}{\sqrt{4Dt}}\right) - \text{erf}\left(\frac{3\Delta}{\sqrt{4Dt}}\right) + \text{erf}\left(\frac{5\Delta}{\sqrt{4Dt}}\right) - \cdots\right)$$

Then first term is a probability density to leave through interfaces. Through one interface we get:

$$f^*(t) = \frac{\Delta}{\sqrt{4\pi Dt^3}} \exp\left(-\beta t - \frac{\Delta^2}{4Dt}\right) - \frac{3\Delta}{\sqrt{4\pi Dt^3}} \exp\left(-\beta t - \frac{9\Delta^2}{4Dt}\right) + \frac{5\Delta}{\sqrt{4\pi Dt^3}} \exp\left(-\beta t - \frac{25\Delta^2}{4Dt}\right) - \cdots$$

And after the integration over whole time:

$$F^* = \exp(-\Delta\sqrt{\beta/D}) - \exp(-3\Delta\sqrt{\beta/D}) + \exp(-5\Delta\sqrt{\beta/D}) - \cdots = \frac{\exp(-\Delta\sqrt{\beta/D})}{\exp(-2\Delta\sqrt{\beta/D}) + 1}$$

Which is smaller than $F$, since photons can leak through the other interface in this case. For thick samples or strong scattering, the second interface does not play a role and $F^*$ reduces to $F$ (denominator approaches unity). The formula for $F^*$ can be expressed through the hyperbolic secant: $F^* = \text{sech}(\Delta\sqrt{\beta/D})/2$. Then the fractions of reflected, transmitted, and absorbed light are ($\Delta = d/2$):

$$F_R = F_T = F^* = \text{sech}(\Delta\sqrt{\beta/D})/2$$

$$F_A = 1 - 2F^* = 1 - \text{sech}(\Delta\sqrt{\beta/D})$$

### S2.4 Diffusion model for the finite slab (source at an arbitrary position)

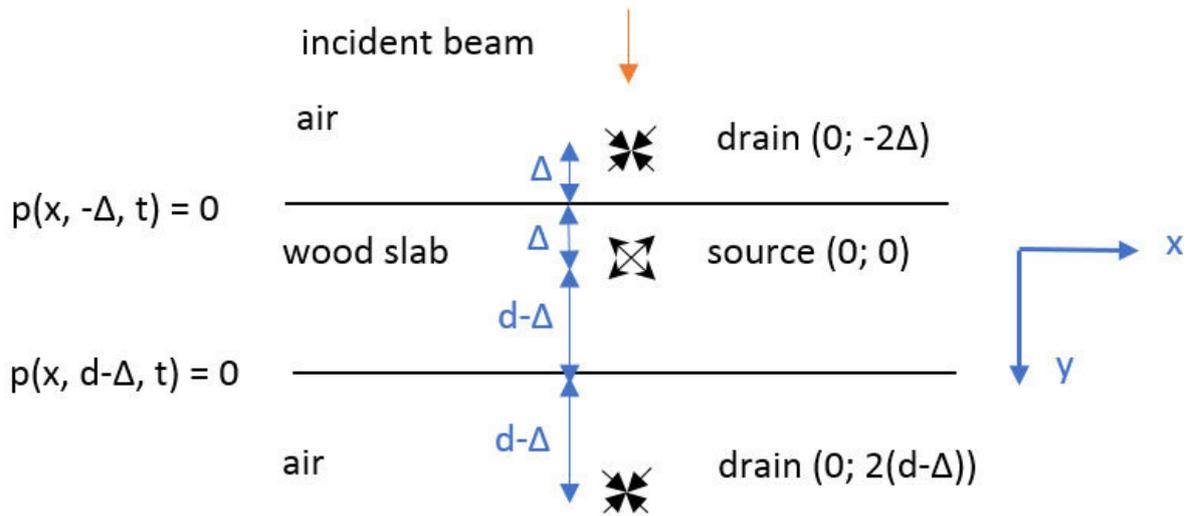

Now consider the point source at a position $\Delta$ from the top interface. The solution:

$$p^{**}(x,y,t) = \frac{\exp\left(-\frac{x^2+y^2}{4Dt}\right) - \exp\left(-\frac{x^2+(y-2\Delta)^2}{4Dt}\right) - \exp\left(-\frac{x^2+(y+2(d-\Delta))^2}{4Dt}\right) + \cdots}{4\pi Dt} \exp(-\beta t)$$

Flux through the output interface:

$$|f_+(x,t)| = D\frac{\partial}{\partial y}p(x,y,t)|_{y=d-\Delta} = \frac{d-\Delta}{4\pi Dt^2}\exp\left(-\beta t - \frac{x^2+(d-\Delta)^2}{4Dt}\right) - \frac{d+\Delta}{4\pi Dt^2}\exp\left(-\beta t - \frac{x^2+(d+\Delta)^2}{4Dt}\right) + \cdots$$

Integrated over $x$-coordinate:

$$f_+(t) = \frac{d-\Delta}{\sqrt{4\pi Dt^3}}\exp\left(-\beta t - \frac{(d-\Delta)^2}{4Dt}\right) - \frac{d+\Delta}{\sqrt{4\pi Dt^3}}\exp\left(-\beta t - \frac{(d+\Delta)^2}{4Dt}\right) + \frac{3d-\Delta}{\sqrt{4\pi Dt^3}}\exp\left(-\beta t - \frac{(3d-\Delta)^2}{4Dt}\right) - \cdots$$

And after the integration over the whole time:

$$F_+ = e^{-(d-\Delta)\sqrt{\beta/D}} - e^{-(d+\Delta)\sqrt{\beta/D}} + e^{-(3d-\Delta)\sqrt{\beta/D}} - e^{-(3d+\Delta)\sqrt{\beta/D}} + \cdots$$

$$F_+ = \frac{\exp\big((d+\Delta)\sqrt{\beta/D}\big) - \exp\big((d-\Delta)\sqrt{\beta/D}\big)}{\exp\big(2d\sqrt{\beta/D}\big) - 1} = \frac{\sinh(\Delta\sqrt{\beta/D})}{\sinh(d\sqrt{\beta/D})}$$

Similarly, back-flux through the input interface

$$F_- = \frac{\sinh((d-\Delta)\sqrt{\beta/D})}{\sinh(d\sqrt{\beta/D})}$$

For $d = 2\Delta$ the result coincides with the previous case. Finally, the absorbed fraction:

$$F_A^* = 1 - F_+ - F_- = 1 - \frac{\cosh((d/2-\Delta)\sqrt{\beta/D})}{\cosh(d\sqrt{\beta/D}/2)}$$

These results can be also obtained through Laplace transform solutions of the diffusion-decay equation.[S5,S6]

### S2.5 Diffusion model for the finite slab (distributed source)

In a real situation the source of diffused photons is not a point, but a probabilistically spread distribution. Probability density to Rayleigh scatter at a point $y$ for the incoming photons, which are not specular reflected and not absorbed in the slab before $y$ (can be either forward scattered or not):

$$\mu_{sr}(1-SR)\exp(-\mu_{sr}y) \cdot \int_y^\infty \mu_a \exp(-\mu_a \xi)\, d\xi = \mu_{sr}(1-SR)\exp(-(\mu_{sr}+\mu_a)y)$$

Then Rayleigh diffused transmittance, reflectance, and absorption:

$$DT_r = \mu_{sr}(1-SR)\int_0^d \exp(-(\mu_{sr}+\mu_a)y)\frac{\sinh(\alpha y)}{\sinh(\alpha d)}dy =$$

$$= \frac{\mu_{sr}(1-SR)\big((\mu_{sr}+\mu_a+\alpha)e^{(\alpha-\mu_{sr}-\mu_a)d} - (\mu_{sr}+\mu_a-\alpha)e^{-(\alpha+\mu_{sr}+\mu_a)d} - 2\alpha\big)}{2\sinh(\alpha d)(\alpha^2 - (\mu_{sr}+\mu_a)^2)}$$

$$DR_r = \mu_{sr}(1-SR)\int_0^d \exp(-(\mu_{sr}+\mu_a)y)\frac{\sinh(\alpha(d-y))}{\sinh(\alpha d)}dy =$$

$$= \frac{\mu_{sr}(1-SR)\big((\mu_{sr}+\mu_a+\alpha)e^{-\alpha d} - (\mu_{sr}+\mu_a-\alpha)e^{\alpha d} - 2\alpha e^{-(\mu_{sr}+\mu_a)d}\big)}{2\sinh(\alpha d)(\alpha^2 - (\mu_{sr}+\mu_a)^2)}$$

$$A_r = \mu_{sr}(1-SR)\int_0^d \exp(-(\mu_{sr}+\mu_a)y)\left(1 - \frac{\cosh(\alpha(d/2-y))}{\cosh(\alpha d/2)}\right)dy =$$

$$= \frac{\mu_{sr}(1-SR)(1-\exp(-(\mu_{sr}+\mu_a)d))}{(\mu_{sr}+\mu_a)} - DT_r - DR_r$$

### S2.6 Comparison to experiment

We can distinguish several fractions after light passes a TW sample. First, non-interacting with the sample are the specular reflection $SR$ from the top interface ($SR = 4\%$), and ballistic photons $BT$:

$$BT = (1-SR)\cdot \exp(-(\mu_{sr}+\mu_{sf}+\mu_a)d) \qquad (1)$$

For samples with a strong scattering and low absorption the interacted part can be separated into the random walk (described by diffusion theory) fraction, and forward scattered fraction.

Absorbed photons on the first passage without Rayleigh scattering (with and without forward scattering – the same optical paths $d$):

$$A_f = \frac{\mu_a(1 - SR)\,(1 - \exp(-(\mu_{sr} + \mu_a)d))}{(\mu_{sr} + \mu_a)}$$

Diffusely transmitted photons with forward scattering and without absorption or Rayleigh scattering:

$$DT_f = (1 - SR)\big(1 - \exp(-\mu_{sf}d)\big)\exp(-(\mu_{sr} + \mu_a)d) = (1 - SR)\exp(-(\mu_{sr} + \mu_a)d) - BT$$

Due to highly directional nature of forward scattering there is no backscattering from those $DR_f \approx 0$.

Total quantities:

$$A = A_r + A_f \qquad (2)$$

$$DR \approx DR_r \qquad (3)$$

$$DT = DT_r + DT_f \qquad (4)$$

Total transmittance:

$$TT = BT + DT = BT + DT_r + DT_f = DT_r + (1 - SR)\exp(-(\mu_{sr} + \mu_a)d)$$

Quantities $DR$, $A$, $DT$, and $BT$ are measurable (the first three with an integrating sphere and the latter with a bucket detector). So, one can get $\mu_{sr}, \mu_a, \mu_{sf}$.

One can numerically solve these equations for samples with a measured photon budget.